\documentclass[onecolumn,showpacs,preprintnumbers,amsmath,amssymb]{revtex4-2}
\usepackage{amsmath, amssymb}
\usepackage{natbib}
\usepackage{graphicx}
\usepackage{setspace}
\usepackage{gensymb}
\usepackage{dcolumn}
\usepackage{bm}
\usepackage[usenames,dvipsnames]{color}
\usepackage{xcolor}
\usepackage{bm}
\def\bk{\bf k}
\newcommand{\be}{\begin{equation}}
\newcommand{\ee}{\end{equation}}
\newcommand{\beq}{\begin{eqnarray}}
\newcommand{\eeq}{\end{eqnarray}}
\begin{document}
\title{Variance of the Casimir force in an ideal Bose gas}
\author{Marek Napi\'{o}rkowski and Marcin Pruszczyk}
 \affiliation{Institute of Theoretical Physics, Faculty of Physics \\ University of Warsaw, Pasteura 5, 02-093 Warsaw, Poland}
\date{\today}

\begin{abstract}
We consider an ideal Bose gas enclosed in a $d$-dimensional slab of thickness $D$. Using the grand canonical ensemble we calculate the variance of the thermal  Casimir force acting on the slab's walls. The variance evaluated per unit wall area  is shown to decay like  $ \Delta_{var}/D$ for large $D$. The amplitude   $ \Delta_{var}$  is a  non-universal function of  two scaling variables $\lambda/ \xi$ and $D/ \xi$, where $\lambda$ is the thermal de Broglie wavelength and $\xi$ is the bulk correlation length. It can be expressed via the bulk pressure, the Casimir force per unit wall area, and its derivative with respect to chemical potential. For thermodynamic states corresponding to the presence of  the Bose-Einstein condensate  the amplitude $ \Delta_{var}$  retains its non-universal character  while the ratio of the mean standard deviation and the Casimir force takes the scaling form $(D/L)^{\frac{d-1}{2}}\, (D/\lambda)^{d/2}$, where $L$ is the linear size of the wall. 
\end{abstract}
\maketitle
\newpage
\section{Introduction}
Fluctuations taking place in finite systems are influenced by the bounding walls. In the case of electromagnetic vacuum within a slab with metallic walls this influence is manifested by the paradigmatic electromagnetic Casimir force \cite{Cas1,Barton1991} . Similarly, thermal fluctuations of material systems enclosed in a slab result in thermal Casimir force \cite{Mostepanienko_97,Krech_94,Kardar_99,Bordag_01,BAFG2002,KD2004,HHGDB2009,HHG2008,Gambassi_09,MZ2006,GD2006,NP1,Napiorkowski_12,NJN1,DiehlRut2017,DS2022}. For a given material system the thermal Casimir force  evaluated per unit area of the walls depends on the distance $D$ between the walls,  the thermodynamic state of the system, and on the boundary conditions.  It becomes particularly conspicuous when the thermodynamic state of the system,  e.g., parametrized by  temperature $T$ and  chemical potential $\mu$,  approaches its bulk critical state $(T_{c}, \mu_{c})$. In this particular state the Casimir force $f_{C}( T_{c},\mu_{c},D)$ becomes long-ranged, i.e., it displays the power law decay for large $D$. The leading order contribution to $f_{C}$ takes  the universal form 
$f_{C}(T_{c},\mu_{c},D) / k_{B}T_{c} = \Delta_{C}/D^d$, where $d$ is the dimensionality of the system and $\Delta_{C}$ denotes the universal amplitude whose numerical value depends on the type of boundary conditions imposed at the walls. In the vicinity of the critical state, in the scaling regime where both the system's bulk correlation length $\xi$ and the distance $D$ tend to infinity with their ratio $D/\xi$ finite, the amplitude $\Delta_{C}$ is  a universal function of $D/\xi$ and the leading order contribution to $f_{C}(T,\mu,D)$ takes the form $f_{C}(T,\mu,D)/k_{B}T =  \Delta_{C}(D/\xi)/D^d$. \\ 
Systematic evaluation of the Casimir force for a given many-body system enclosed in a slab with volume $L^{d-1}D$ requires the calculation of the system's total free energy including, inter alia,  its bulk and surface contributions. In this paper we work in the grand canonical ensemble and the grand canonical potential is 
\beq
\label{Om1}
\Omega(T,\mu, L,D) = L^{d-1} D\,\omega_{b}(T,\mu) + L^{d-1}\,\omega_{s}(T,\mu, D) + \dots ,
\eeq
where $\omega_{b}(T,\mu)$ denotes the bulk free energy density and $\omega_{s}(T,\mu, D)$ is the surface free energy density. While the analysis of 
$\omega_{b}(T,\mu)$ allows one to determine the system's bulk properties including its phase diagram and the bulk critical properties, the surface contribution 
$\omega_{s}(T,\mu, D)$ gives the Casimir force per unit area 
\beq
\label{cas201}
f_{C}(T,\mu,D) =  -  \frac{\partial \omega_{s}(T,\mu,D)}{\partial D} \,= \, - \lim\limits_{L \rightarrow \infty} \frac{1}{L^{d-1}}\, \frac{\partial \Omega(T,\mu, L,D)}{\partial D} - p(T,\mu) ,
\eeq
where $p = - \omega_{b}$ is the bulk pressure. \\

The fluctuations-induced Casimir force, Eq.(\ref{cas201}), can be also evaluated as the ensemble average of appropriately chosen stochastic variable ${\cal F}_{C}$ with the corresponding probability distribution $\cal{P}(\cal{F}_{C})$.  In this paper we analyze a particular characteristic of this  probability distribution, namely its variance. \\
Our discussion concentrates on an ideal  Bose gas of $N$ spinless bosons enlosed in a $d$-dimensional slab $L^{d-1} D$.  From the point of view of rigorous analysis this is a particularly convenient system. On one hand it  displays a phase transition to a phase which hosts the Bose-Einstein condensate. On the other hand both the Casimir force and the variance can be calculated analytically.  The system's Hamiltonian is 
\beq
 {\hat{{\cal{H}}}} = \sum_{\bf k} \frac{\hbar^2{\bf k}^2}{2m}\,\hat{n}_{\bf k} ,
\eeq
where $m$ denotes the boson's mass and we use the standard second-quantization notation. The system is subject to Neumann  
boundary conditions which imply that ${\bf k} = (k_{1}, \dots, k_{d}) = (\frac{\pi n_1}{L},\dots,\frac{\pi n_{d-1}}{L},\frac{\pi n_d}{D},)$ with $n_{i}=0,1,2,\dots$, $i=1,\dots,d$. In the case of Dirichlet b.c. the analogous analysis becomes much more involved due to the non-zero single particle ground state energy, see e.g. \cite{NPT2020}, and is postponed to further analysis. \\
The Casimir force which in Eq.(\ref{cas201}) is specified as the $D$-derivative of the grand canonical free energy per unit wall area, can be equivalently evaluated as the grand canonical average of 
$ - \frac{\partial {\hat{\cal{H}}}}{\partial D}$ 
\beq
\label{Cas10}
f_{C}(T,\mu,D) = \lim\limits_{L \rightarrow \infty} \,\frac{1}{L^{d-1}}\, \left<  - \frac{\partial {\hat{\cal{H}}}}{\partial D} \right> - p 
\eeq 
corrected for the contribution stemming from the bulk pressure $p$. 
Indeed, it is straightforward to check that 
\beq
\lim\limits_{L \rightarrow \infty} \, \frac{1}{L^{d-1}}\,\left<  \frac{\partial {\hat{\cal{H}}}}{\partial D} \right>  =  \lim\limits_{L \rightarrow \infty} \, \frac{1}{L^{d-1}}\,\frac{\partial \Omega(T,\mu, L,D)}{\partial D} .
\eeq
We note that 
\beq
\label{clas01}
\frac{1}{L^{d-1}}\,\left(-  \frac{\partial {\hat{\cal{H}}}}{\partial D} \right) = \sum\limits_{n_{1},\dots,n_{d}} \,{\hat{n}}_{n_{1},\dots,n_{d}}\,\frac{\hbar\pi\,n_{d}}{D} \,\frac{\hbar \pi\,n_{d}}{D} \,\frac{1}{m}\,\frac{1}{L^{d-1}D} \, = \,  \sum\limits_{n_{1},\dots,n_{d}} \, p_{z\,n_{d}}\frac{\,p_{z\,n_{d}}}{m} \frac{{\hat{n}}_{n_{1},\dots,n_{d}}}{L^{d-1}D}
\eeq
and thus the expressions  in Eqs(\ref{Cas10},\ref{clas01}) represent the quantum counterpart of classical expressions for evaluating the force acting on the container's wall. It is  based on the mechanical definition of pressure which employs averaging of the momentum current density \cite{LL,Mishin,HS2018}. 
At the same time the procedure of evaluating the Casimir force $f_{C}$ specified in Eq.(\ref{Cas10}) suggests a straightforward way to calculate the variance 
\beq
\label{flc01}
\sigma_{f_{C}}^{2} = \lim\limits_{L \rightarrow \infty} \, \frac{1}{L^{d-1}}\,\left(\left<\left(\frac{\partial  {\hat{\cal{H}}}}{\partial D}\right)^2\right> - 
\left<\frac{\partial  {\hat{\cal{H}}}}{\partial D}\right>^{2} \right) .
\eeq
The variance $\sigma_{f_{C}}^{2} $ provides additional information to the one already contained in the average value $f_{C}$. We shall evaluate the expression for the variance and discuss its scaling properties, in particular its $D$-dependence,  upon approaching the Bose-Einstein condensation. \\

The above expression closely resemble the one for the  bulk pressure $p$ evaluated as ensemble average \cite{LL,ZKU}. In this case one considers a $d$-dimensional cube $\underbrace {L\times \dots \times L}_{d}$ and 
\beq
\label{p01}
p = \,\,\lim\limits_{V \rightarrow \infty}  \left<  - \frac{\partial {\hat{\cal{H}}}}{\partial V} \right>  \, = \,\lim\limits_{L \rightarrow \infty} \,\frac{1}{d\,L^{d-1}}\, 
\left<  - \frac{\partial {\hat{\cal{H}}}}{\partial L} \right> .
\eeq
In the case of the Casimir force one considers anisotropic container $\underbrace{L \times \dots \times L}_{d-1} \times D$, takes the limit $L \rightarrow \infty$ first and then considers the large $D$-behaviour of the resulting expression. \\

\section{Variance of the Casimir force}
Before calculating the variance of the Casimir force we recall the corresponding results for the bulk pressure $p$ of an ideal Bose gas \cite{LL,ZKU}. We restrict our considerations to $d > 2$ to allow for the existence of the Bose-Einstein condensation. For ideal Bose gas the space of thermodynamic states $(T,\mu)$ is restricted to non-positive $\mu$-values. It follows from Eq.(\ref{p01}) that for $\mu < 0 $, which corresponds to the absence of the condensate,  
\beq
\label{p02}
p(T,\mu)  = \,\lim\limits_{L \rightarrow \infty} \,\frac{1}{dL^{d-1}}\, \left<  - \frac{\partial \cal{H}}{\partial L} \right>  = \frac{k_{B}T}{\lambda^d}\,
g_{\frac{d+2}{2}}\left(e^{\beta\mu}\right) .
\eeq
where $g_{\kappa}(z) = \sum\limits_{m=1}^{\infty} \frac{z^m}{m^\kappa} $ is the Bose function,  $\lambda = \frac{h}{\sqrt{2\pi m k_BT}}$ is  the thermal de Broglie wavelength, and $1/\beta = k_{B}T$. The bulk density $n_b$ is (for $\mu < 0$) 
\beq
\label{p03}
n_b = \,\,\lim\limits_{L \rightarrow \infty} \frac{<N>}{L^d} =  \frac{1}{\lambda^{d}} \,g_{\frac{d}{2}}\left(e^{\beta\mu}\right) .
\eeq
If instead of $(T,\mu)$ one considers $(T,n_b)$ as independent thermodynamic variables then the expression for the pressure $p$ in Eq.(\ref{p02})  holds  for 
$n_b \leq n_{c}(T) = \lambda^{-d}\,\zeta(d/2)$, where $\zeta(\kappa) = g_{\kappa}(1)$ denotes the Riemann zeta function. On the other hand, for $\mu= 0$ and $T < T_{c}$ , or equivalently for $n_{b}  >  n_{c}(T)$, one has 
\beq
\label{p002}
p(T,0)  = \frac{k_{B}T}{\lambda^d} \, \zeta\left(\frac{d+2}{2}\right) 
\eeq
and $ n_b = n_{c}(T) + n_{0}(T)$, where $n_{0}(T)$ is the condensate density. 
The variance $\sigma^{2}_{p} $ characterizing the fluctuations of pressure 
\beq
\sigma^{2}_{p} =  \left<  \left(- \frac{\partial {\hat{\cal{H}}}}{\partial V}  \right)^2 \right> -  \left<  - \frac{\partial {\hat{\cal{H}}}}{\partial V} \right>^2
\eeq
 can be straightforwardly transformed and evaluated as
 \beq
 \label{p07}
 \sigma^{2}_{p} = k_{B}T \, \left( \left< \frac{\partial^2{{\hat{\cal{H}}}}}{\partial V^2} \right> - \frac{\partial^2 \Omega}{\partial V^2}  \right)\, = \, \frac{d+2}{d} 
 \,\frac{k_{B}T}{V}\,p \,. 
\eeq
 Identical result  for the variance $\sigma^{2}_{p} $ is obtained from the thermodynamic theory of fluctuations \cite{LL,Mishin,ZKU,HS2018} which predicts 
 $\sigma^{2}_{p} = - k_{B}T \,\left( \partial p/\partial V\right)_{S,N}$. When this expression is supplemented by the expression for the bulk entropy density 
  $s_{b}(T,\mu)  =  \left(\frac{d+2}{2}\,p(T,\mu) - \mu\,n(T,\mu) \right)/T$ \cite{ZKU}, and transformed with the help of  thermodynamic identities then one recovers the result in Eq.(\ref{p07}). \\
 
After these introductory remarks we evaluate the variance of the Casimir force in the grand canonical ensemble,  Eq.(\ref{flc01}).  First we observe that
\beq
\left<\left(\frac{\partial  {\hat{\cal{H}}}}{\partial D}\right)^2 \right> 
= \frac{1}{\beta} \left<\frac{\partial^2{\hat{\cal{H}}}}{\partial D^2} \right> \, + \, \frac{1}{\beta^2 \,\Xi(T,\mu,L,D)}\,\frac{\partial^2\Xi(T,\mu,L,D)}{\partial D^2} ,
\eeq
where $\Xi(T,\mu,L,D)$ is the grand canonical partition function. With the help of 
\beq
\left<\frac{\partial  {\hat{\cal{H}}}}{\partial D}\right>^{2} \, = \, \frac{1}{\beta^2\,\Xi^2} \, \left(\frac{\partial \Xi}{\partial D}\right)^2
\eeq
one straightforwardly obtains 
\beq
\label{flc02}
\left<\left(\frac{\partial  {\hat{\cal{H}}}}{\partial D}\right)^2\right> - \left<\frac{\partial  {\hat{\cal{H}}}}{\partial D}\right>^{2} \, = \, \frac{1}{\beta} \, 
\left[\left<\frac{\partial^2{\hat{\cal{H}}}}{\partial D^2} \right> \, - \,\frac{\partial^2 \Omega}{\partial D^2} \right]. 
\eeq

For the ideal Bose gas one has $\frac{\partial^2{{\hat{\cal{H}}}}}{\partial D^2} = - \frac{3}{D}\frac{\partial{{\hat{\cal{H}}}}}{\partial D} $ and the variance in Eq.(\ref{flc01}) can be rewritten in the following form 
\beq
\label{flc07}
\sigma_{f_{C}}^{2} =  \frac{1}{\beta} \, \left[ \left(\frac{\partial f_{C}}{\partial D}\right)_{T,\mu} \, + \,\lim\limits_{L \rightarrow \infty} \, \frac{1}{L^{d-1}}\, \left<\frac{\partial^2{{\hat{\cal{H}}}}}{\partial D^2} \right>\right] = \frac{1}{\beta}\,\frac{1}{D^3}\left.\frac{\partial}{\partial D}\right|_{T,\mu}\, \left[D^3\,(f_{C}+p)\right] \,.
\eeq

The grand canonical potential $\Omega(T,\mu,L,D) $ is 
\beq
\label{grcan}
\Omega(T,\mu,L,D) = - k_{B}T \log \Xi(T,\mu,L,D)  = k_{B}T\,\sum\limits_{n_{1}=0}^{\infty}\, \dots \sum\limits_{n_{d}=0}^{\infty}\, 
\log\left(1- z e^{-\frac{\pi\lambda^2}{4}\left(\frac{n_1^2}{L^2} + \dots + \frac{n_{d-1}^2}{L^2} + \frac{n_d^2}{D^2} \right)}\right) , 
\eeq 
where $z=e^{\beta\mu}$. In order to evaluate the rhs of Eq.(\ref{flc02}) we rewrite the first term as 
\beq
\label{flc03}
\left<\frac{\partial^2{\hat{\cal{H}}}}{\partial D^2} \right>  = \left< \sum\limits_{\bk} {\hat{n}}_{\bk} \,\frac{\partial^2 \epsilon_{\bk}}{\partial D^2} \right> 
= \frac{3 \hbar^2 \pi^2}{m D^4} \, \left< \,\sum\limits_{n_{1}=0}^{\infty}\, \dots \sum\limits_{n_{d}=0}^{\infty}\,{\hat{n}}_{n_1,\dots,\,n_{d}}\,n_{d}^2\right> = \nonumber \\
= \frac{3 \hbar^2 \pi^2}{m D^4} \, \,\sum\limits_{n_{1}=0}^{\infty}\, \dots\, \sum\limits_{n_{d}=0}^{\infty} 
\frac{n_{d}^2}{e^{\frac{\pi\lambda^2}{4}\left(\frac{n_1^2}{L^2} + \dots + \frac{n_{d-1}^2}{L^2} + \frac{n_d^2}{D^2} \right)}\,z^{-1} -1} \, .
\eeq
We note that both terms on the rhs of Eq.(\ref{flc02}), i.e., $\left<\frac{\partial^2{\hat{\cal{H}}}}{\partial D^2} \right>$ in Eq.(\ref{flc03}) and $\frac{\partial^2 \Omega}{\partial D^2}$ in Eq.(\ref{grcan}),  display similar structure which can be concisely represented as 
$\sum\limits_{n_{1}=0}^{\infty}\, \dots\, \sum\limits_{n_{d}=0}^{\infty} \, \varphi(n_1,\dots,n_d) $. The form 
of the function $\varphi(n_1,\dots,n_d) $ depends on the particular macroscopic quantity to be evaluated, i.e., it is either 
$ \log\left(1-  e^{\beta \mu - \frac{\pi\lambda^2}{4}\left(\frac{n_{1}^2}{L^2} + \dots + \frac{n_{d-1}^2}{L^2} + \frac{n_{d}^2}{D^2} \right)}\right)$ or 
$\frac{n_{d}^2}{e^{\frac{\pi\lambda^2}{4}\left( \frac{n_{1}^2}{L^2} + \dots + \frac{n_{d-1}^2}{L^2} + \frac{n_{d}^2}{D^2} \right) - \beta \mu} -1} $.  \\
When analyzing the expressions  in  Eqs(\ref{grcan},\ref{flc03})  for 
$L \rightarrow \infty$ and finite $D$ it is convenient to use the following formula \cite{Napiorkowski_12} 
\beq
\label{sum001}
\sum\limits_{n=0}^{\infty} \psi(n) = \frac{1}{2}\psi(0) + \int\limits_{0}^{\infty} \,dx\, \psi(x) + 2 \sum_{m=1}^{\infty} \int\limits_{0}^{\infty} dx\, 
\psi(x) \cos(2\pi m x) .
\eeq
It follows from the Poisson summation formula for an even function $\psi(x)$ such that $\psi(\infty)=0$; see also \cite{ZKU,Saharian2007}.  The application of this formula  to the sum $\sum\limits_{n_{1}=0}^{\infty}\, \dots\, \sum\limits_{n_{d}=0}^{\infty} \, \varphi(n_1, \dots,n_d) $ allows to extract the contributions $\sim L^{d-1}$ to each term on the rhs of Eq.(\ref{flc02}), and then analyze their $D$-dependence. The details of the calculations are postponed to the Appendix. \\

\noindent Below we discuss discuss both the Casimir force, Eq.(\ref{Cas10}) and its variance, Eqs(\ref{flc01},\ref{flc02},\ref{grcan},\ref{flc03}) using the results obtained in the Appendix. 
One can straightforwardly check that the expression for the variance can be rewritten in the following simple form
\beq
\label{flc012} 
\sigma_{f_{C}}^{2} = \lim\limits_{L \rightarrow \infty} \, \frac{1}{L^{d-1}}\,\left(\left<\left(\frac{\partial  {\hat{\cal{H}}}}{\partial D}\right)^2\right> - 
\left<\frac{\partial  {\hat{\cal{H}}}}{\partial D}\right>^{2} \right) = 
\left(\frac{\pi\lambda^{\frac{5-d}{2}}}{2\beta}\right)^2\,\frac{1}{D^6} \sum\limits_{n_{d}=0}\,n_{d}^4\,\,g_{\frac{d-3}{2}}\left(e^{\beta\mu - \frac{\pi\lambda^2}{4D^2} n_{d}^2}\right)\,.
\eeq
The expression for the Casimir force takes the form  
\beq
\label{Casd01}
\frac{f_{C}(T,\mu;D) }{k_{B}T} \, =  \,  \frac{\Delta_{C}\left(\frac{D}{\xi}\right)}{D^d} , 
\eeq
where 
\beq 
\label{Cas200d}
\Delta_{C}(x) = \,2^{\frac{4-3d}{2}}\,\pi^{-\frac{d}{2}}\,x^{\frac{d}{2}}\,\sum\limits_{p=1}
\frac{K_{\frac{d}{2}}\left(x p\right) - x p \,K_{\frac{d}{2}+1} \left(x p\right) }{p^{\frac{d}{2}}} 
\eeq
is the Casimir force amplitude. The symbol $K_{\nu}(z)$ denotes the modified Bessel function of second kind of order $\nu$. The amplitude $\Delta_{C}(x)$ depends on the distance $D$ and the thermodynamic state $(T,\mu)$ via the ratio $D/\xi$,  
where $\xi = \frac{\lambda}{\sqrt{16 \pi \beta (-\mu)}} $ is the bulk correlation length \cite{ZKU,Napiorkowski_12}. The Casimir force is attractive. Upon approaching the Bose-Einstein condensation at $\mu=0$, the correlation length $\xi \rightarrow \infty$ and the Casimir force amplitude  $ \Delta_{C}(x)$  tends to the universal value
\beq
\label{Cas200d0}
\Delta_{C}(0) \, = \,2^{1-d}\,\pi^{-\frac{d}{2}}\,\Gamma \left(\frac{d}{2}\right)\,\zeta(d)\,(1- d)\,.
\eeq
For $d=3$ one recovers the previous result $\Delta_{C}(0) = - \zeta(3)/4\pi$,  \cite{MZ2006}. 
The amplitude $\Delta_{C}(0)$ is a negative, non-monotonous function of $d$ with a maximum near $d=24$.  For large $d$ one has 
$\Delta_{C}(0) \sim -  d^{(d+1)/2}/(8 \pi e)^{d/2}$.   \\ 
On the other hand, the variance $\sigma_{f_{C}}^{2}$  takes the following form 
\beq
\label{var100d}
\frac{\sigma_{f_{C}}^{2}(T,\mu,D)}{(k_{B}T)^2}   = \, \frac{\Delta_{var}(T,\mu,D)}{D} 
\eeq
with the non-universal amplitude $\Delta_{var}(T,\mu,D)$ given by 
\beq
\label{var103}
\lambda^d\, \Delta_{var}(T,\mu,D)  = 3 \,g_{\frac{d+2}{2}}\left(e^{-\frac{1}{16\pi}\,\left(\frac{\lambda}{\xi}\right)^2}\right) + \left(\frac{\lambda}{\xi}\right)^{d}\,\Psi\left(\frac{D}{\xi}\right) = 
\frac{3\,\lambda^d\,p(T,\mu)}{k_{B}T} + \left(\frac{\lambda}{\xi}\right)^{d}\,\Psi\left(\frac{D}{\xi}\right) \,,
\eeq
where
\beq
\label{ampd200}
\Psi(x) = \frac{4}{(8\pi)^{\frac{d}{2}}}\,\sum\limits_{p=1}\,\frac{3\,K_{\frac{d}{2}}(x\,p) - 6\,x\,p\,K_{\frac{d}{2}+1}(xp) + x^2\,p^2\,K_{\frac{d}{2}+2}(x\,p)}{(x\,p)^{\frac{d}{2}}} \,.
\eeq  
This result clearly displays the dependence of  $\lambda^d\, \Delta_{var}(T,\mu,D) $ on two scaling variables $D/\xi$ and $\lambda/\xi$. \\ 
For $d=3$ our results reduce to  
\beq
\label{varamp01}
\lambda^3\, \Delta_{var}^{(3)}(T,\mu,D) \, =  \, \left[ 3 g_{\frac{5}{2}}\left(e^{-\frac{1}{16\pi}\,\left(\frac{\lambda}{\xi}\right)^2}\right) + 
\frac{1}{8\pi}\, \left(\frac{\lambda}{\xi}\right)^3\, \frac{1}{e^{\frac{D}{\xi}} - 1} \right] \,.
\eeq 
It follows that for arbitrary $d$ and for large $D/\lambda$, the dominant contribution to the variance decays $\sim 1/D$ . In particular, for $\mu=0$ the 
second term on the rhs of Eq.(\ref{var103}) decays as 
\beq
\lim\limits_{\mu \rightarrow 0} \, \left(\frac{\lambda}{\xi}\right)^{d}\,\Psi\left(\frac{D}{\xi}\right) = \frac{2^{1-d}}{\pi^{\frac{d}{2}}} \, \Gamma\left(\frac{d}{2}\right)\,\zeta(d)(1-d)(3-d)\,\left(\frac{\lambda}{D}\right)^d = (3-d)\,\Delta_{C}(0)\,\left(\frac{\lambda}{D}\right)^d  \,.
\eeq
Thus in the condensed phase and for large $D$ the dominant contribution to the variance takes  the following form 
 \beq
\frac{\sigma_{f_{C}}^{2}(T,0,D)}{(k_{B}T)^2}   = \, \frac{3\,p(T,0)}{k_{B}T}\,\frac{1}{D} \,.
\eeq
We note that the $\sim 1/D$ decay law of the variance upon increasing $D$  is different than the one of the Casimir force itself as specified in Eqs(\ref{Casd01},\ref{Cas200d},\ref{Cas200d0}).  In the non-condensed phase, i.e., for $\mu < 0$,  the leading contribution to the variance  decays like $1/D$ while for the Casimir force it has the form $e^{-D/\xi}/D^d$. In the condensed phase the decays of the variance and the Casimir force are different as well: the variance still decays like $1/D$ while the Casimir force decays like $1/D^d$. Thus the large $D$-decay of the variance is always slower than that of the Casimir force. 
The above result describing the large $D$-decay of the variance differs from the predictions of analysis in \cite{BAFG2002,KD2004}, where different models, namely the scalar field model with Dirichlet boundary conditions and lattice spin models were discussed, respectively. There,  the obtained expression for the variance of the Casimir force led to the conclusion that it  does not decay to zero upon increasing $D$.

The expression for the variance $\sigma_{f_{C}}^{2}$ in Eq.(\ref{flc07}) can be further simplified with the help of the scaling property of the Casimir force. 
Indeed, the Casimir force in Eqs(\ref{cas201},\ref{Cas10}) displays the following scaling 
\beq
f_{C}(\Lambda^2\,T, \Lambda^2\,\mu,\Lambda^{-1}\,D) = \Lambda^{d+2}\,f_{C}(T,\mu,D) \, ,
\eeq
where $\Lambda$ is the scaling parameter, and thus 
\beq
\label{scfc03}
D\,\frac{\partial f_{C}(T,\mu,D)}{\partial D} = \left(- (d+2) + 2\mu\frac{\partial}{\partial \mu} + 2 T\frac{\partial}{\partial T} \right)\,f_{C}(T,\mu,D)\,. 
\eeq
After using the equality $\left(\frac{\partial \frac{f_{C}}{T}}{\partial T}\right)_{\mu,D} =0$, see Eqs(\ref{Casd01},\ref{Cas200d}), this leads to the following expression for the amplitude $\Delta_{var}(T,\mu,D) $ in Eq.(\ref{var100d}) 
\beq
\label{Delta02}
\Delta_{var}(T,\mu,D) = \frac{3\,p(T,\mu) + (3-d)f_{C}(T,\mu,D) + 2\mu \frac{\partial f_{C}(T,\mu,D)}{\partial \mu}}{k_{B}T} \,.
\eeq
This expression can be rewritten using the thermodynamic relation for the slab 
\beq
\label{difom01}
\left. d \omega_{s}(T,\mu,D)\right|_{T} = - f_{C}\, dD - 2\,n_{s}\, d\mu 
\eeq
and 
\beq
\label{surterm}
\left(\frac{\partial f_{C}(T,\mu,D)}{\partial \mu}\right)_{T,D} = 2\,\left(\frac{\partial n_{s}(T,\mu,D)}{\partial D}\right)_{T,\mu} \,,
\eeq
where $n_{s}$ is the surface density of particles. The quantity $n_{s}$  can be extracted from the following expression 
\beq
\label{surden01d}
\lim\limits_{L \rightarrow \infty} \,\frac{1}{L^{d-1}}\, \left<  N \right>  = D\,n_{b} + 2\,n_{s}  \,,
\eeq
where the bulk density $n_{b}$ is given in Eq.(\ref{p03}) and 
\beq
\label{surden02d}
n_{s}(T,\mu,D) = \frac{1}{4\,\lambda^{d-1}} \,g_{\frac{d-1}{2}}\left(e^{\beta\mu}\right) + \frac{2^{4-\frac{3d}{2}}\,\pi^{1-\frac{d}{2}}}{\lambda^{d-1}} \left(\frac{D}{\xi}\right)^{\frac{4-d}{2}}\,
\left(\frac{\lambda}{\xi}\right)^{d-3} \sum\limits_{p=1}\frac{K_{\frac{d-2}{2}}\left(\frac{D}{\xi}p\right)}{p^{\frac{d-2}{2}}} \,.
\eeq
With the help of Eqs(\ref{Casd01},\ref{Cas200d},\ref{surden02d}) and the recurrence relations for Bessel functions  one can directly verify the thermodynamic relation in Eq.(\ref{surterm}). It follows that 
\beq
\label{Delta03}
\Delta_{var}(T,\mu,D) = \frac{3\,p(T,\mu) + (3 - d)\,f_{C}(T,\mu,D)}{k_{B}T} \,- 
\, \frac{1}{4\pi}\,\left(\frac{\lambda}{\xi}\right)^2\,\left(\frac{\partial n_{s}(T,\mu,D)}{\partial D}\right)_{T,\mu}\,.
\eeq
Thus the amplitude of the variance can be expressed via the bulk pressure, the Casimir force, the $D$-derivative of the surface density, and the ratio of the bulk correlation length and the thermal de Broglie wavelength. 
We note that in the expression on the rhs of Eq.(\ref{Delta03}) the contribution to the amplitude $\Delta_{var}$ stemming from the Casimir force itself vanishes for $d=3$. The leading contribution to the amplitude  $\Delta_{var}$ for large $D$ is proportional to the bulk pressure, and the variance decays $\sim 1/D$. \\
If one considers the expressions for the Casimir force and the variance on the rhs of  Eqs(\ref{Cas10},\ref{flc01}) for large but finite values of $L$, 
 $L/\lambda \gg 1$, and takes the dominant contributions to the total Casimir force to be 
 $ F_{C} = L^{d-1} f_{C}$ and similarly for the variance $\sigma_{F_{C}}^{2} = L^{d-1}\, \sigma_{f_{C}}^{2}$, then the ratio of the standard deviation $\sigma_{F_{C}}$ and the total Casimir force $F_{C}$ evaluated in the condensed phase at $\mu=0$ takes the scaling form 
\beq
\label{ratio01}
\frac{\sigma_{F_{C}} }{ F_{C}} \, \sim  \, \left(\frac{D}{L}\right)^{\frac{d-1}{2}}\,\left(\frac{D}{\lambda}\right)^{\frac{d}{2}} \, . 
\eeq
The above result differs from the predictions obtained in \cite{BAFG2002,KD2004} for scalar field model and lattice spin models (in our analysis the thermal de Broglie wavelength $\lambda$ plays the role of a microscopic length scale in \cite{BAFG2002}). The difference pertains to the value of one of the exponents characterizing this scaling behavior.  Namely, the exponent $\frac{d}{2}$ in Eq.(\ref{ratio01})  replaces the exponent $\frac{d+1}{2}$ in \cite{BAFG2002,KD2004}. The reason for this discrepancy is that in our analysis of the Bose gas the variance $\sigma_{f_{C}}^{2}$ is shown to decay  $\sim 1/D$ while for systems considered in \cite{BAFG2002,KD2004} it remains $D$-independent. 

\section{Summary}
We analyzed the  ideal Bose gas enclosed in a $d$-dimensional slab of thickness $D$ with Neumann boundary conditions imposed at the slab walls of area $L^{d-1}$. We have calculated the Casimir force and its variance as the grand canonical averages. In particular we discussed the behavior of these two quantities as function of the Bose gas thermodynamic state $(T, \mu)$ and the distance between the walls $D$ in the limit $L \rightarrow \infty$. For large distances $D$ we recovered the known results for the Casimir force amplitude \cite{MZ2006,GD2006,Napiorkowski_12} and its scaling properties. In the case of the variance we showed that independently of the system dimensionality $d$ it decays for large $D$  as $1/D$. This is a much slower decay then the one characterizing the Casimir force itself which decays as $1/D^d$. This prediction remains in contrast with previous studies of different models in which the obtained expression for the variance turned out to be $D$-independent. We have also determined the amplitude characterizing the variance                                   
$\Delta_{var}(T,\mu,D)$ 
in the scaling regime. We have shown that $\lambda^d\,\Delta_{var}(T,\mu,D)$ is a function of two scaling variables $\lambda/\xi$ and $D/\xi$. Note that the corresponding amplitude for the Casimir force depends on a single  scaling variable $D/\xi$. When our results are applied to the finite but large $L$ case, then the ratio of the total variance and the total Casimir force evaluated for the condensed phase  takes the universal scaling form   $ \left(\frac{D}{L}\right)^{\frac{d-1}{2}}\,\left(\frac{D}{\lambda}\right)^{\frac{d}{2}} $.  \\ 
One of the open problems in the context of Casimir force is related to the equivalence of  canonical and grand canonical results as far as the variance is concerned, see \cite{GVGD2016,PCCH2016,GGD2017,RSVDG2019}. It is well known that these two ensembles lead to different results for the ideal Bose gas as far as the density fluctuations in the condensed phase are concerned \cite{ZKU}. The relevant  question whether similar inequivalence holds also in the case of Casimir force fluctuations is left for further analysis.

\section{Appendix}
In this Appendix we present a systematic method  which allows us to extract the bulk and surface contributions to typical quantities relevant for the present analysis. For simplicity of notation we consider  the $d=3$ case. First we note that in each of the two cases under consideration, Eqs(\ref{grcan},\ref{flc03}), the function $\varphi(n_1,n_2,n_3)$ is symmetric under the interchange of the first two arguments $\varphi(n_1,n_2,n_3)= \varphi(n_2,n_1,n_3)$. 
As a result of threefold application of Eq.(\ref{sum001}) one obtains the following representation 
\beq
\label{sum01}
\sum\limits_{n_{1}=0}^{\infty}\, \sum\limits_{n_{2}=0}^{\infty}\, \sum\limits_{n_{3}=0}^{\infty} \, \varphi(n_1,n_2,n_3) = \frac{1}{8} \varphi(0,0,0) + \nonumber \\
+ \frac{1}{2}\, \int\limits_{0}^{\infty} dx\,\varphi(x,0,0) \left[1+2 \sum_{p=1}^{\infty}\,\cos(2\pi p x)\right] + \frac{1}{4}\, \int\limits_{0}^{\infty} \,dx\,\varphi(0,0,z) \left[1+2 \sum_{r=1}^{\infty}\,\cos(2\pi r z)\right] + \nonumber \\
+ \frac{1}{2}\, \int\limits_{0}^{\infty} dx\int\limits_{0}^{\infty} dy \varphi(x,y,0) \left[1+4 \sum_{p=1}^{\infty}\,\cos(2\pi p x) + 
4 \sum_{p=1}^{\infty}\,\sum_{q=1}^{\infty}\,\cos(2\pi p x)\,\cos(2\pi q y) \right] +\nonumber \\
+ \int\limits_{0}^{\infty} dx \int\limits_{0}^{\infty} dz \varphi(x,0,z) \left[1+2 \sum_{p=1}^{\infty}\,\cos(2\pi p x) + 2 \sum_{r=1}^{\infty}\,\cos(2\pi r z) +
4 \sum_{p=1}^{\infty}\,\sum_{r=1}^{\infty}\,\cos(2\pi p x)\,\cos(2\pi r z) \right] + \nonumber \\
+ \int\limits_{0}^{\infty} dx\int\limits_{0}^{\infty} dy \int\limits_{0}^{\infty} dz \varphi(x,y,z)\,\left[1+ 4 \sum_{p=1}^{\infty}\,\cos(2\pi p x) + 2 \sum_{r=1}^{\infty}\,\cos(2\pi r z) + 4 \sum_{p=1}^{\infty}\,\sum_{q=1}^{\infty}\,\cos(2\pi p x)\,\cos(2\pi q y) + \right. \nonumber \\ 
\left. + 8 \sum_{p=1}^{\infty}\,\sum_{r=1}^{\infty}\,\cos(2\pi p x)\,\cos(2\pi r z) + 8 \sum_{p=1}^{\infty}\,\sum_{q=1}^{\infty}\,\sum_{r=1}^{\infty}\,\cos(2\pi p x)\,\cos(2\pi q y)\,\cos(2\pi r z) \right]\,. 
\eeq
In each case under consideration the functions $\varphi(x,0,0)$, $\varphi(x,y,0)$, 
$\varphi(x,0,z)$, and $\varphi(x,y,z)$ in Eq.(\ref{sum01}) depend on the integration variables $x,y,z$  via the rescaled variables $x/L, y/L, z/D$. In the case of grand canonical potential $\Omega(T,\mu,L,D)$ it follows from Eq.(\ref{sum01}) that 
\beq
\label{app01}
\lim\limits_{L \rightarrow \infty} \frac{\Omega(T,\mu,L,D)}{L^2} = - \frac{k_{B}T}{\lambda^2}\,\sum\limits_{n_{3}=0}^{\infty}\,g_{2}\left(e^{\beta \mu - \frac{\pi\lambda^2}{4D^2}\,n_{3}^2}\right) = \nonumber \\ 
= - \frac{k_{B}T}{\lambda^2}\,\left[  \frac{1}{2}\,g_{2}\left(e^{\beta\mu}\right) \, + \, \sum\limits_{n=1}^{\infty} \frac{e^{\beta \mu n}}{n^2}\,\int\limits_{0}^{\infty}dz 
e^{- \frac{\pi\lambda^2 z^2 n}{4 D^2}} \, + \, 2 \sum\limits_{r=1}^{\infty} \sum\limits_{n=1}^{\infty} \frac{e^{\beta\mu n}}{n^2} \, \int\limits_{0}^{\infty}dz 
\cos(2\pi r z)\,e^{- \frac{\pi \lambda^2 z^2 n}{4 D^2}}\right] = \nonumber \\
= - k_{B}T\,\left[ \frac{1}{2\lambda^2}\,g_{2}\left(e^{\beta\mu}\right) + \frac{D}{\lambda^3}\,g_{\frac{5}{2}}\left(e^{\beta\mu}\right) + \frac{1}{8\pi}\frac{1}{D^2} 
 \sum\limits_{r=1}^{\infty} \, e^{-\frac{D}{\xi}r}\left(\frac{1}{r^3} + \frac{D}{\xi\,r^2}\right)\right] 
\eeq
and one obtains
\beq
\label{app02}
\lim\limits_{L \rightarrow \infty} \frac{1}{L^2}\, \frac{\partial^2\Omega(T,\mu,L,D)}{\partial D^2} = -\frac{k_{B}T}{8\pi D^4}\left[
6 g_{3}\left(e^{-\frac{D}{\xi}}\right) + 6 \frac{D}{\xi} g_{2}\left(e^{-\frac{D}{\xi}}\right) + 3 \left(\frac{D}{\xi}\right)^2 g_{1}\left(e^{-\frac{D}{\xi}}\right) + 
\left(\frac{D}{\xi}\right)^3 g_{0}\left(e^{-\frac{D}{\xi}}\right) \right].
\eeq
The analysis of the remaining term contributing to the variance follows along similar  lines
\beq
\label{app03}
\lim\limits_{L \rightarrow \infty} \frac{1}{L^2}\,\left<\frac{\partial^2\cal{H}}{\partial D^2} \right>  = \lim\limits_{L \rightarrow \infty} \frac{1}{L^2}\,\frac{3 \hbar^2 \pi^2}{m D^4} \, \,\sum\limits_{n_{1}=0}^{\infty}\, \sum\limits_{n_{2}=0}^{\infty}\, \sum\limits_{n_{3}=0}^{\infty} 
\frac{n_{3}^2}{e^{\frac{\pi\lambda^2}{4}\left(\frac{n_1^2}{L^2} + \frac{n_2^2}{L^2} + \frac{n_3^2}{D^2} \right)}\,z^{-1} -1}  = \nonumber \\
= \frac{48}{\pi^{\frac{3}{2}}}\frac{k_{B}T}{D\lambda^3} \left[ \int\limits_{0}^{\infty}dx \int\limits_{0}^{\infty}dy \int\limits_{0}^{\infty}dz 
\frac{z^2}{e^{x^2+y^2+z^2-\beta\mu}-1} + 2\sum\limits_{r=1}^{\infty}\,\int\limits_{0}^{\infty}dx \int\limits_{0}^{\infty}dy \int\limits_{0}^{\infty}dz 
\frac{z^2}{e^{x^2+y^2+z^2-\beta\mu}-1}\,\cos\left(4\pi \frac{D}{\lambda}r z\right) \right] = \nonumber \\
=    \frac{48}{\pi^{\frac{3}{2}}}\frac{k_{B}T}{D\,\lambda^3} \left[  \frac{\pi^{\frac{3}{2}}}{16}\,g_{\frac{5}{2}}\left(e^{\beta\mu}\right) \, + \, \frac{\pi}{2} 
\sum\limits_{r=1}^{\infty}\,\sum\limits_{n=1}^{\infty}\,\frac{e^{\beta \mu n}}{n^{\frac{5}{2}}}\,\int\limits_{0}^{\infty} dz z^2\,e^{-z^2}\,\cos\left(4\pi^{\frac{1}{2}}\frac{D}{\lambda} \frac{r}{n^{\frac{1}{2}}}z \right) \right] \, = \nonumber \\
= \frac{48}{\pi^{\frac{3}{2}}}\frac{k_{B}T}{D\,\lambda^3} \left[  \frac{\pi^{\frac{3}{2}}}{16}\,g_{\frac{5}{2}}\left(e^{\beta\mu}\right) \,- \, 
\frac{\pi^{\frac{3}{2}}}{16} \sum\limits_{r=1}^{\infty}\,\sum\limits_{n=1}^{\infty}\,\,\frac{e^{\beta \mu n}}{n^{\frac{5}{2}}}\, \, 
\left(16\pi\frac{D^2}{\lambda^2}\frac{r^2}{n} - 2 \right)\,e^{-4\pi\frac{D^2}{\lambda^2}\frac{r^2}{n}} \right]\, = \nonumber \\
= \frac{3\,k_{B}T}{D\,\lambda^3} \left\{ g_{\frac{5}{2}}\left(e^{\beta\mu}\right) - \frac{1}{8\pi}\left(\frac{\lambda}{\xi}\right)^3
\left[ 2 \left(\frac{\xi}{D}\right)^3g_{3}\left(e^{-\frac{D}{\xi}}\right) + 2 \left(\frac{\xi}{D}\right)^2 g_{2}\left(e^{-\frac{D}{\xi}}\right) \, + \, 
\frac{\xi}{D}\, g_{1}\left(e^{-\frac{D}{\xi}}\right) \right] \right\} .
\eeq
Combination of the results corresponding to Eqs(\ref{app02},\ref{app03})  for the $d$-dimensional case gives the final result in Eqs(\ref{var103},\ref{ampd200}).

\begin{acknowledgements} 
We thank Jaros{\l}aw Piasecki and  Pawe{\l} Jakubczyk for their valuable comments and suggestions concerning the manuscript. 
\end{acknowledgements}

\end{document}